# Nitridation of InP(100) surface studied by synchrotron radiation

M. Petit[a], D. Baca[a,b], S. Arabasz[c], L. Bideux[a], N. Tsud[d], S. Fabik[d], B. Gruzza[a], V. Chab[e], V. Matolin[b], K.C. Prince[d]

a. LASMEA, UMR 6602 CNRS, Blaise Pascal University, Campus scientifique des Cézeaux, 63177 Aubiere Cedex
b. Department of Electronics and Vacuum Physics, Charles University, V Holesovickach 2, 18000 Prague 8, Czech Republic
c. Department of Microelectronics, Institute of Physics, Silesian University of Technology, Krzywoustego 2, 44-100 Gliwice, Poland
d. Sincrotone Trieste, Strada Statale 14, km 163.5, 34012 Trieste, Italy
e. Institute of Physics, Czech Academy of Science, Cukronvarnicka 10, 16200 Prague, Czech Republic

Corresponding author : Luc BIDEUX
Fax : + 33 4 73 40 73 40
e-mail : bideux@lasmea.univ-bpclermont.fr

**Abstract:**

The nitridation of InP(100) surfaces has been studied using synchrotron radiation photoemission. The samples were chemically cleaned and then ion bombarded, which cleaned the surface and also induced the formation of metallic indium droplets. The nitridation with a Glow Discharge Cell (GDS) produced indium nitride by reaction with these indium clusters. We used the In 4d and P 2p core levels to monitor the chemical state of the surface and the coverage of the species present. We observed the creation of In-N and P-N bonds while the In-In metallic bonds decrease which confirm the reaction between indium clusters and nitrogen species. A theoretical model based on stacked layers allows to assert that almost two monolayers of indium nitride are produced. The effect of annealing on the nitridated layers at 450°C has been also analysed. It appears that this system is stable until this temperature, well above the congruent evaporation temperature (370°C) of clean InP(100) : no increase of metallic indium bonds due to decomposition of the substrate is detected as shown in previous works [16] studying the InP(100) surfaces

# I. Introduction

III-V nitride semiconductors are very promising materials for metal-insulator-semiconductor (MIS), optoelectronic devices and solar cells [1]. The nitridation step often plays a key role in the development of these devices. It is useful for the formation of thin nitridated layers used as passivating films, the deposition of an insulating film on GaAs or InP MIS diodes and for the creation of buffer layers for epitaxial growth of thicker nitridated films [2].

However the nitridation of III-V semiconductors is quite difficult because of the small sticking probability of nitrogen on the surface [3]. For InP, several kinds of nitridation process have been used to produce nitridated films: among them, alkali metals promoted nitridation which lead to the formation of mainly $InPN_x$ layers [4,5], direct nitrogen ion beam nitridation resulting in nitridated layers showing In-N, In-N-P and P-N bondings [3] and plasma nitridation which entails the creation of In-N and P-N bonds [2].

In this article we report the nitridation of InP(100) surfaces with a glow discharge cell. The composition of the nitridated layers was studied using synchrotron radiation photoemission. The passivating effect of nitridated layers under annealing is also examined.

# II. Experimental

The InP(100) wafers used were n-type with carrier concentration of $4.7 \times 10^{16}$ cm$^{-3}$. Before the introduction in the ultrahigh vacuum chamber, they were chemically cleaned in different ultrasonic baths (deionised water, $CH_3OH$, $H_2SO_4$ 96%, Br-$CH_3OH$) [6]. After introduction into the chamber, a low amount of carbon and oxygen contamination was detected. These impurities were removed by in-situ cleaning with low energy Ar$^+$ ions (300 eV, sample current density 2μ A.cm$^{-2}$). This ion bombardment cleaning is a key step in the nitridation process since it creates at the surface metallic indium droplets in well-controlled quantities (mean coverage : 25%, mean height: 4 atomic monolayers) [7,8].

Photoemission experiments were carried out using the Material Science Beamline (MSB) at the Elettra Synchrotron (Trieste, Italy). It is a bending magnet beamline with a tuning range from 40 to 800 eV. The UHV experimental chamber with a base pressure of $1\times10^{-10}$ mbar is equipped with a 150 mm mean radius electron energy analyser Phoibos150 made by Specs with a multichannel detection. The analyser wes used in a constant energy resolution mode. Core level spectra were recorded at an emission angle of 30° from the surface normal and 30° incidence of photons. $In_{4d}$ and $P_{2p}$ core levels were measured with photon beam energies of hν =50 eV and 190 eV. The binding energy scale and total resolution have been calibrated using the Fermi level of a Au plate.

### III. Results and discussion

#### 1. Cleaning of the substrates

Figure 1 shows the evolution of the $In_{4d}$ core level at these two. We observed some differences between the two spectra, particularly there is a contribution (A) at 16,7 eV binding energy clearly visible at 50 eV. This contribution can be attributed to metallic indium [9] due to the indium droplets created by the ionic cleaning. The fact that the metallic indium droplets are clearly seen at 50 eV (32 eV kinetic energy) but not at 190 eV (172 eV kinetic energy, less surface sensitive) proves that the formation of indium clusters is a surface phenomenon.

To decompose the $In_{4d}$ and the $P_{2p}$ core levels, we have to know the different chemical environments of the indium and phosphorus atoms. Previous works[7,8] showed that after ionic cleaning, the surface is covered by 25% of metallic indium droplets with an equivalent height of 4 monolayers.

Regarding previous studies on InP(100)[10,11,12,13,14], it is very difficult to find a definitive result on the decomposition of In4d peak, most of the authors found a bulk In-P contribution, and two or three surface peaks assigned to indium adatoms.

The scheme in figure 2 presents a model of such a surface: this is a simple layered model where the InP(100) is schematised by alternate indium and phosphorus layers.

Regarding the indium atoms, four chemical environments can be found: the bulk (environment 1), indium atoms situated at the interface under the droplets (2), indium atoms inside the droplets (3) and indium atoms at the top of droplets (4). So the $In_{4d}$ core level can be decomposed into four contributions corresponding to the four environments. The spectra were fitted using a Shirley background and a decomposition into Gaussian and Lorentzian line shapes. The fitting parameters are summarised in table 1.

Moreover for each environment, we can calculate the theoretical signal coming from the indium atoms:

$$I_1 = \left( 0.75 \frac{\alpha}{1-\alpha^2} + 0.25 \frac{\alpha^5}{1-\alpha^2} \right) i_{In}$$

$$I_2 = \left( 0.25 \alpha^3 \right) i_{In}$$

$$I_3 = \left( 0.25 \alpha + 0.25 \alpha^2 \right) i_{In}$$

$$I_4 = 0,25 \, i_{In}$$

where $i_{In}$ is the photoelectron intensity of one atomic monolayer of indium and $\alpha = \exp\left( \frac{d}{b\lambda_i} \right)$ the attenuation coefficient of $In_{4d}$ photoelectrons by one monolayer where d is the thickness layer, b the apparatus factor and $\lambda_i$ the inelastic mean free path of the photoelectrons.

To determine numerical values of these intensities, we have to know $\alpha$. We used the decomposition of an $In_{4d}$ core level of a cleaned InP(110) [15]. The ratio between bulk and surface contributions is equal to one. Using the same kind of theoretical model as that in figure 2, the ratio of the bulk intensity versus the surface intensity is written : $\frac{I_{bulk}}{I_{surface}} = \frac{\alpha}{1-\alpha}$ which gives a value of 0.5 for $\alpha$ at 50 eV with $\lambda_i$=4,5 Å . Thus, the numerical values of intensities are: $I_1 = 0.51\ i_{In}$, $I_2 = 0.03\ i_{In}$, $I_3 = 0.19\ i_{In}$, $I_4 = 0.25\ i_{In}$.

Figure 3 presents the $In_{4d}$ core level decomposed into three doublets (due to the spin-orbit splitting) corresponding to the different chemical environments. The doublet related to environment 2 is not seen which is consistent with the low theoretical value of $I_2$. The experimental ratio $\frac{I_3 + I_4}{I_1}$ is equal to 0.84, very close to the theoretical value 0.85, which validates the model and confirms the presence of indium metallic droplets at the surface. We note that no oxide components were detected.

Figure 4 shows the $P_{2p}$ core level doublet decomposed into two doublets related to the chemical environment described in the figure 2.

2. Nitridation of the InP(100) surfaces

The indium droplets created by the ion bombardment cleaning are consumed by nitrogen species coming from a glow discharge cell (GDS) to create nitride layers. The GDS dissociates nitrogen molecules using a high voltage (2200 volts). The InP(100) substrate was heated to 250°C, and the sample was kept under a nitrogen flow for 40 min. The temperature and times used were found to be the optimal values for the process in previous works [16-17]. The $In_{4d}$ and $P_{2p}$ core levels are displayed in the figure 5 and 6. On the $In_{4d}$ core level, we observed the disappearance of the component A related to the metallic indium and appearance of two contributions B and C which can be attributed to the formation of nitride layers.

Regarding the $P_{2p}$ core level, a component appears at about 133 eV which is due to the formation of P-N bonds as reported in several papers on the nitridation of InP surfaces [2,3]. P-N bonds are shifted by 4.3 eV from the bulk. We can note that the contribution related to the substrate is also modified, due to the removal of surface P-In bonds (environment 6) since there is a nitride film which covers the InP(100) surface.

The presence of nitrogen atoms on the surface creates new chemical environments for the indium atoms. These are summarised in the figure 7. With this model, the In$_{4d}$ core level after nitridation can be decomposed into three doublets: one for bulk In-P bonds (environment 1), two related to indium nitride (environments 7 and 8) as shown in figure 8. The P$_{2p}$ core level (figure 9) will present a new component corresponding to the P-N bonds (environment 9).

We note the presence of metallic indium in the In$_{4d}$ core level spectrum showing that the indium clusters were not totally consumed by the nitrogen during the nitridation process. In this case, it is also possible to calculate the theoretical photoelectron intensity. We have considered that the coverage rate of the surface by the nitride layers was equal to unity and that there were two layers of stoichiometric indium nitride. Indeed indium droplets represent one complete atomic monolayer of metallic indium ($\theta$ =25%, height = 4 ML). So the maximum number of stoichiometric indium nitride layers which can be formed is two monolayers. The intensity for the bulk In-P is written :

$$I'_1 = \left(\frac{\alpha^3}{1-\alpha^2}\right) i_{In} = 0.166\, i_{In}$$

The intensities coming from In-N bulk and surface bonds (environment 7 and 8) are given by the following expressions:

$$I_7 = (0.5\alpha)\, i_{In} = 0.25\, i_{In}$$

$$I_8 = 0.5\, i_{In}$$

(Numerical values are given with $\alpha$ =0.5)

The experimental ratio $\dfrac{I_7 + I_8}{I'_1}$ is equal to 4.49, close to the theoretical value 4.52. The experimental ratio is a little lower than the theoretical one, which is consistent with the fact that the decomposition of the $In_{4d}$ core level still show the presence of metallic indium, so that the amount of nitride monolayers is lower than two.

### 3. Effect of annealing

We heated the nitridated samples at 450°C. The evolution of the $In_{4d}$ and $P_{2p}$ core level spectra is displayed in figures 5 and 6. Core levels after an annealing at 450°C do not show a significant difference compared to those observed after the nitridation. The relative concentrations of the different bonds are summarised in table 2.
Nitride layers seem to be quite stable during the annealing since no significant change of In-N bonds is observed. To be more precise we observed a small increase of In-N surface bonds which can be due to a 3D-2D transformation of indium nitride at the annealing temperature. Indeed the congruent evaporation temperature of InP is about 370°C: thermal annealing of cleaned InP(100) surfaces has been shown to result in surface decomposition associated with P depletion of the surface and formation of indium droplets [18]. This entails a indium enrichment on the surface. This enrichment is not observed until 450°C in our nitridated InP substrates (see table 2), so it means that the nitride layers block the migration of P atoms until 450°C. InN forms a protective overlayer of the InP surface.

## IV. Conclusions

The nitridation of InP(100) surface with a GDS have been examined using synchrotron radiation photoemission. Ion bombardment creates metallic indium droplets at the surface of the InP(100) sample. These droplets are consumed by the atomic nitrogen flow coming from a GDS to form nitride film. Comparison between a theoretical model based on stacked layers and experimental data shows that almost two monolayers of indium nitride have been formed on the surface. The study of the annealing demonstrates the passivating effect of the nitride layers up to 450°C. There is a very low diffusion of phosphorus atoms in the nitride film and no metallic indium enrichment at the surface sample. InN layers allows thermal stabilization of InP(100) surfaces at higher temperature than the congruent evaporation temperature of cleaned InP.

| Peak | In 4d | | | | | P 2p | | |
|---|---|---|---|---|---|---|---|---|
| Component | Env 1 Bulk In-P | Env. 3 Bulk In-In | Env. 4 Surface In-In | Env. 7 In-N | Env. 8 In-N | Env. 5 Bulk P-In | Env. 6 Surface P-In | Env. 9 P-N |
| Absolute position (eV) | 17.20 ± 0.05 | 16.70 ± 0.05 | 17.60 ± 0.05 | 17.70 ± 0.05 | 18.30 ± 0.05 | 129.05 ± 0.05 | 128.40 ± 0.05 | 133 ± 0.05 |
| Shift relative to bulk In-P (env 1) or P-In (env 5) (eV) | | -0.5 | +0.5 | +1.0 | +1.1 | | -0.6 | +4.0 |
| FWHM (eV) | 0.56 | 0.62 | 0.52 | 0.74 | 0.63 | 0.70 | 0.70 | 0.95 |
| % Lor/Gauss | 13 | 54 | 32 | 27 | 27 | 9 | 9 | 9 |
| Spin-orbit (eV) | 0.86 | 0.86 | 0.86 | 0.86 | 0.86 | 0.9 | 0.9 | 0.9 |
| Branching ratio | 1.45 | 1.45 | 1.45 | 1.45 | 1.45 | 1.95 | 1.95 | 1.95 |

Table 1 : Decomposition parameters of In4d and P2p peaks

|  | Ion bombardment cleaned | After nitridation | After annealing |
|---|---|---|---|
| In-P (environment 1) | 0.54 | 0.17 | 0.17 |
| In-In$_{bulk}$ (environment 3) | 0.18 | 0.03 | 0.03 |
| In-In$_{surface}$ (environment 4) | 0.27 | 0 | 0 |
| In-N (environment 7+8) | 0 | 0.79 | 0.80 |

Table 2 : Variations of the In4d contributions during cleaning, nitridation and annealing of InP(100)

# Figure captions

Figure 1 :     In4d peak of InP(100) after ionic cleaning obtained with a photon beam energy of 50 eV (full line) and 190 eV (dashed line)

Figure 2a:    Side view of the schematic representation of the InP(100) substrate after ionic cleaning.

Figure 2b :   Determination of the chemical environments of indium atoms (dark balls) and phosphorus atoms (light balls) after cleaning. 1 : bulk In-P, 2 : base of In cristallites, 3 : bulk of In cristallites, 4 : top of In cristallites, 5 : bulk P-In, 6 : surface P-In.

Figure 3 :    Experimental In4d peak of cleaned InP(100). This peak can be decomposed in three doublets corresponding to three chemical environments of indium atoms.

Figure 4 :    Experimental P2p peak of cleaned InP(100). This peak can be decomposed in two doublets corresponding to two chemical environments of phosphorus atoms.

Figure 5 :    Experimental In4d peak of InP(100) obtained after cleaning, nitridation and annealing.

Figure 6 :    Experimental P2p peak of InP(100) obtained after cleaning, nitridation and annealing.

Figure 7a :   Side view of the schematic representation of the InP(100) substrate after nitridation.

Figure 7b :   Determination of the chemical environments of indium atoms (dark balls), nitrogen atoms (yellow balls) and phosphorus atoms (light balls)after cleaning. 7 : surface In-N, 8 : bulk In-N, 9 : P-N bonds.

Figure 8 :    Experimental In4d peak of cleaned InP(100). This peak can be decomposed in four doublets corresponding to four chemical environments of indium atoms.

Figure 9 :	Experimental P2p peak of cleaned InP(100). This peak can be decomposed in two doublets corresponding to two chemical environments of phosphorus atoms.

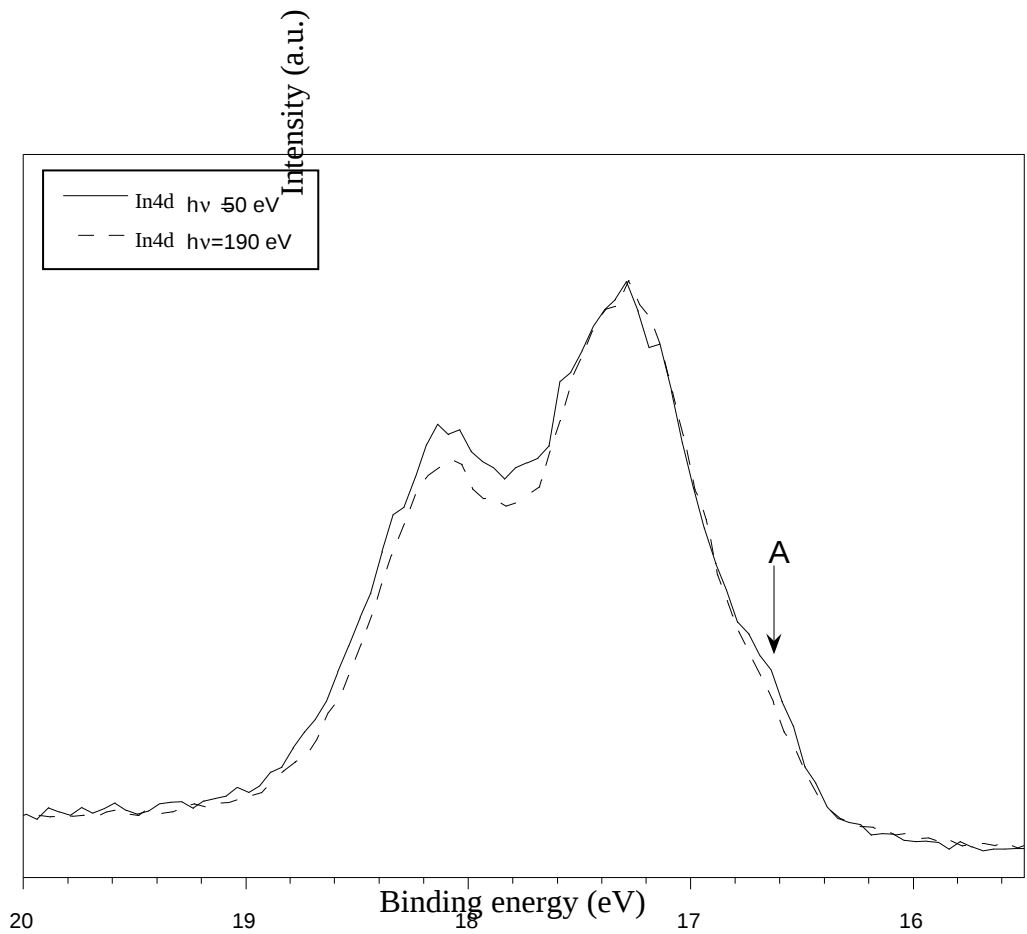

Figure 1

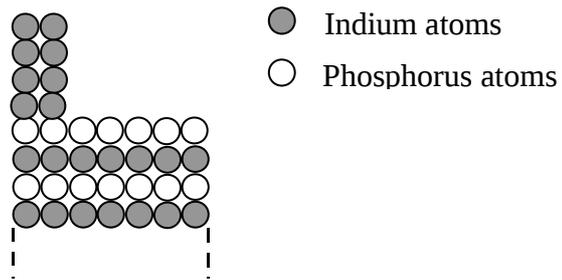

Figure 2a

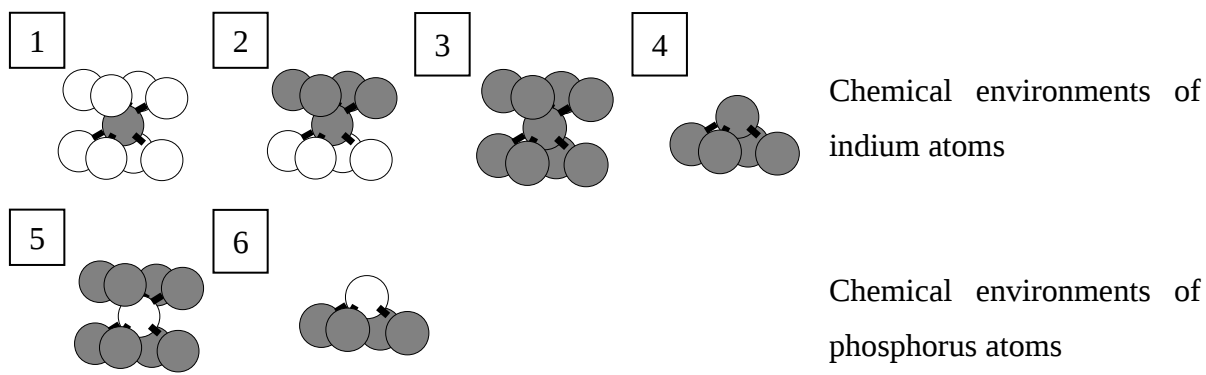

Figure 2b

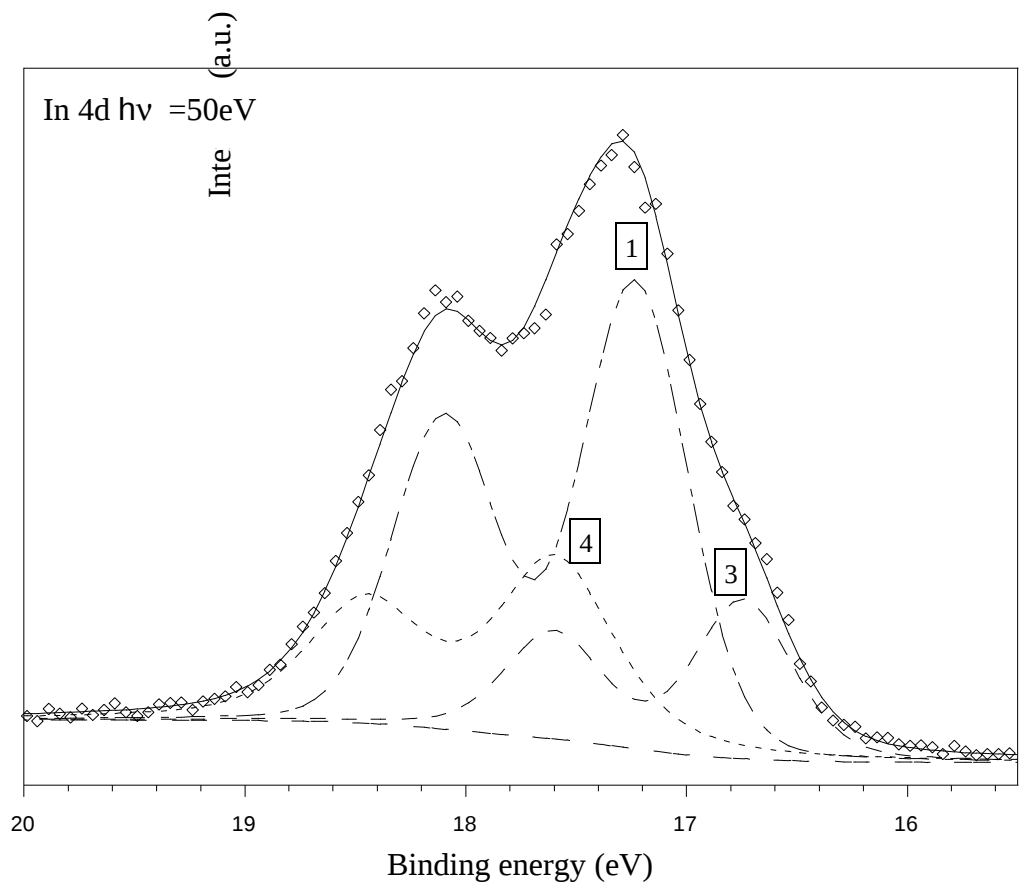

Figure 3

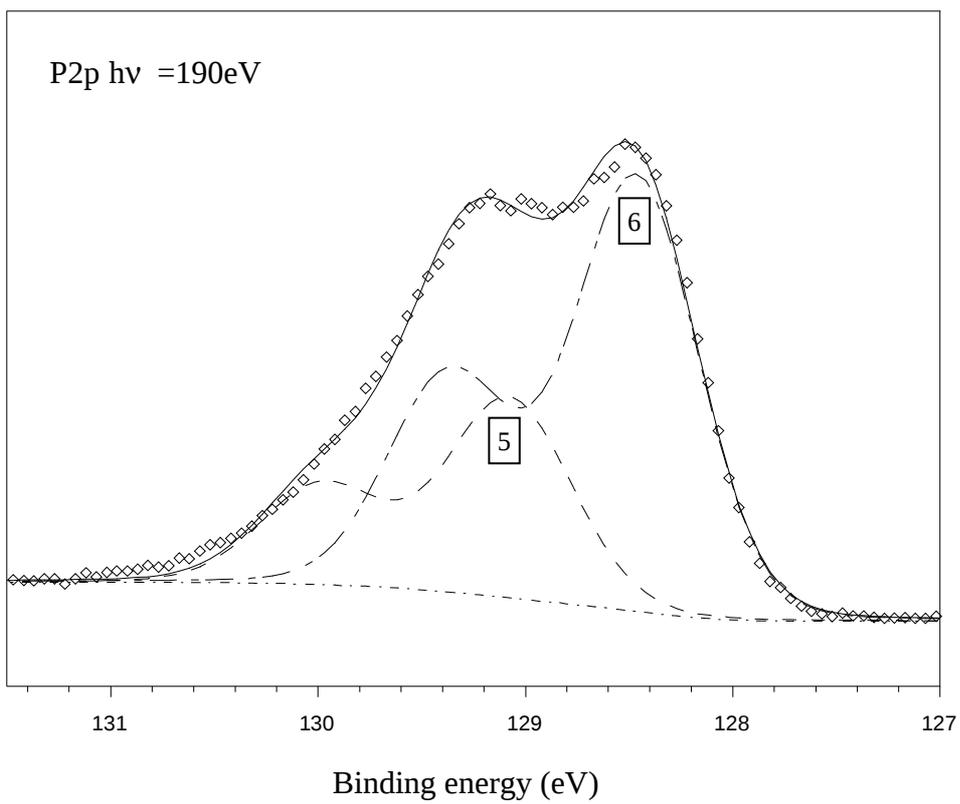

Figure 4

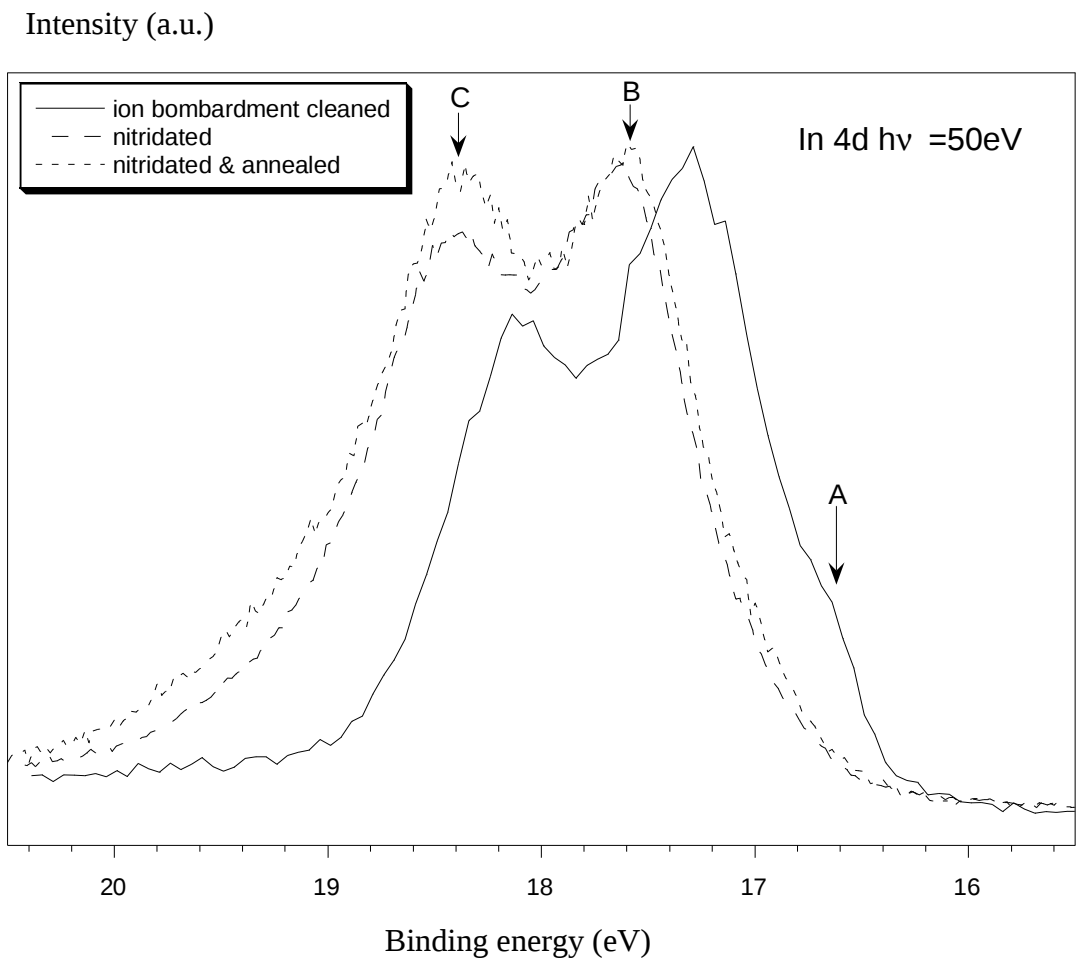

Figure 5

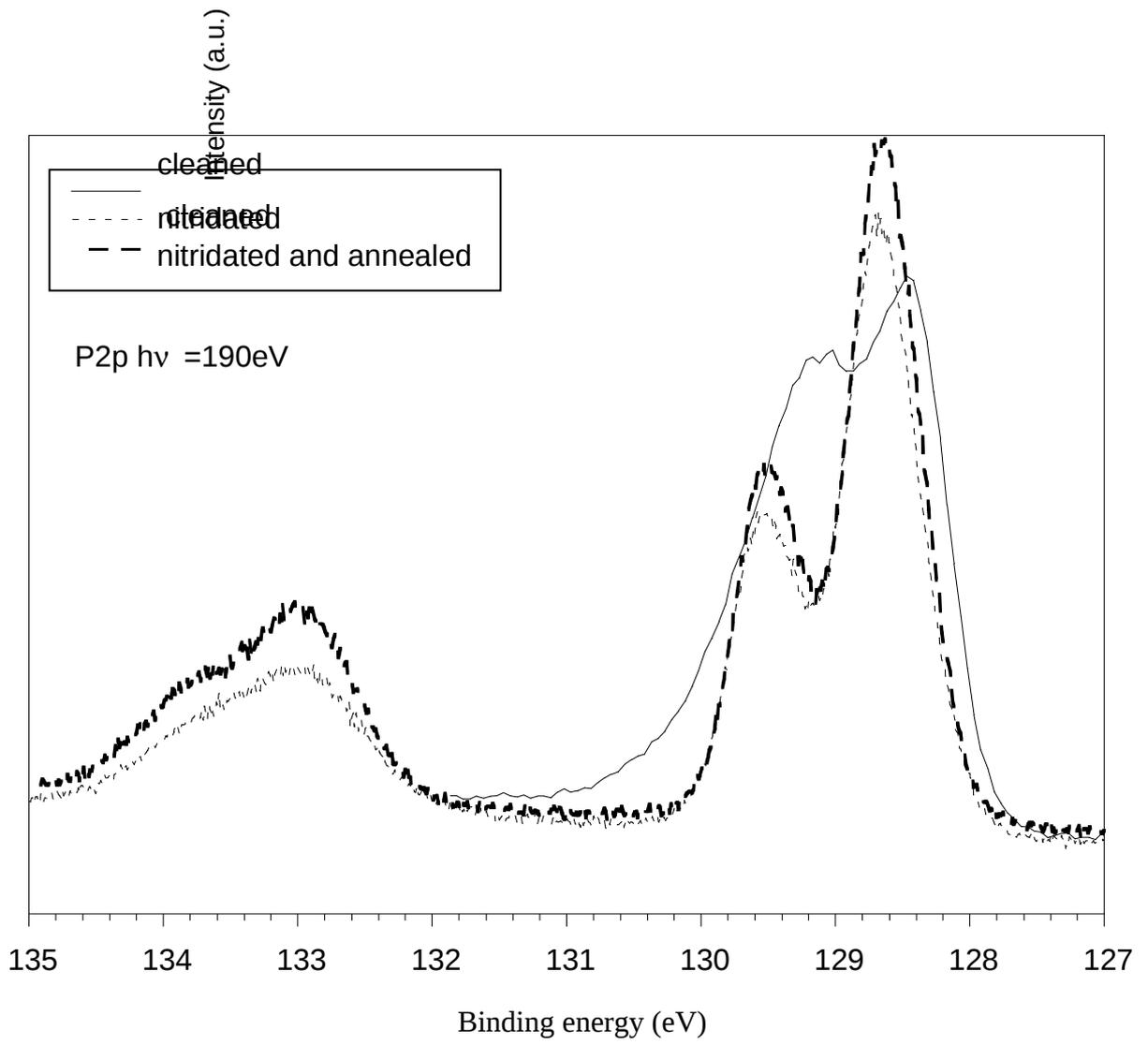

Figure 6

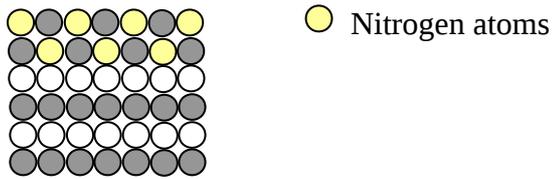 Nitrogen atoms

Figure 7a

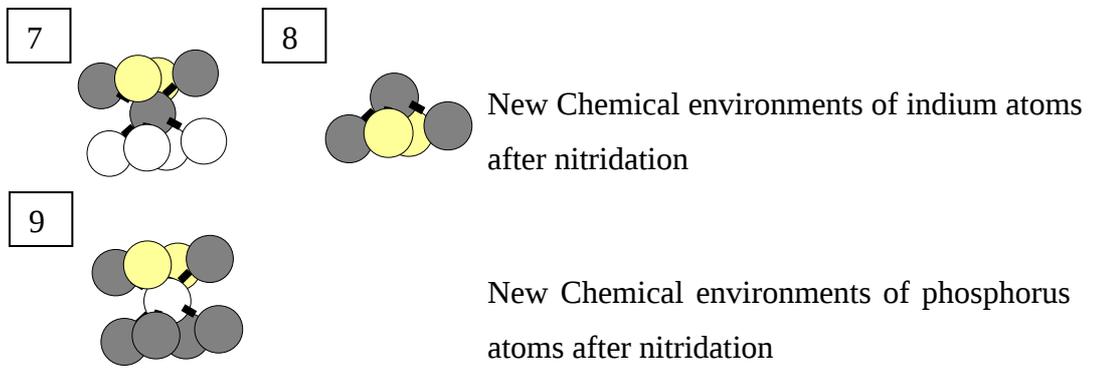

New Chemical environments of indium atoms after nitridation

New Chemical environments of phosphorus atoms after nitridation

Figure 7b

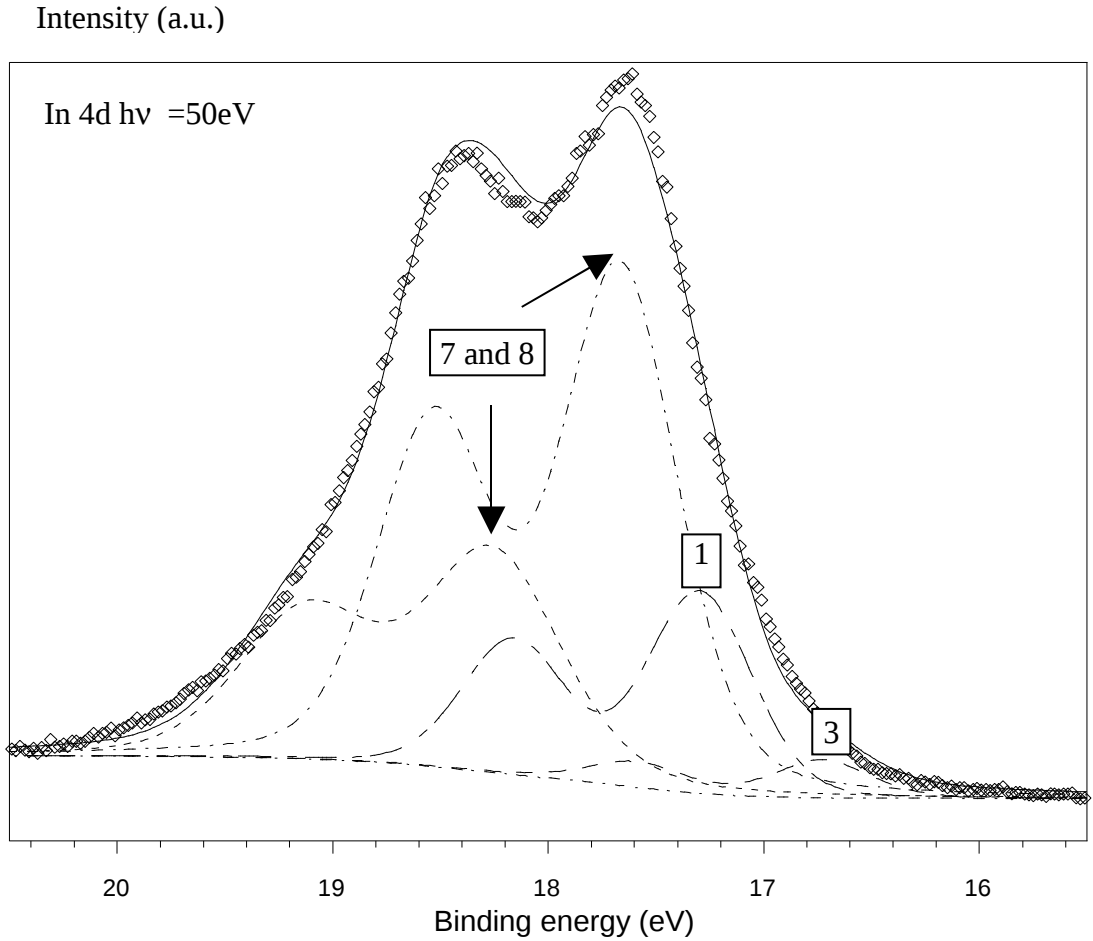

Figure 8

Intensity (a.u.)

P 2p hν =190eV

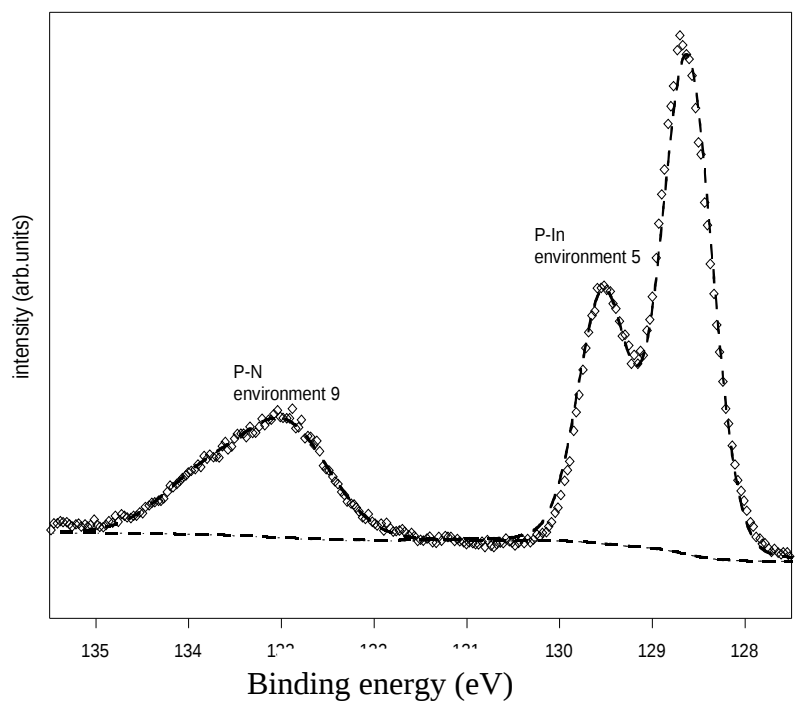

Figure 9